\documentclass[12pt,english,russian]{article}
   \usepackage[cp1251]{inputenc}
   \usepackage{babel}

 \usepackage{graphicx}
\usepackage{grffile}

   \usepackage{amsmath}
\usepackage{amssymb}

\begin{document}

\title
{Detection of 2D electron g-factor sign inversion in narrow GaAs/Al$_x$Ga$_{1-x}$As quantum wells\\}

\author{V.V. Solovyev,$^{1,2}$, W.Dietsche $^{2}$ and I.V. Kukushkin $^{1,2}$ \\
\it (1) Institute of Solid State Physics Russian Academy of Sciences, \\
\it Chernogolovka, Moscow District, 142432 Russia \\
\it (2) Max-Planck-Institute fur Festkorperforschung, 70569
Stuttgart, Germany} \maketitle


\begin{abstract}

By means of CW polarization-resolved photoluminescence spectroscopy we demonstrate the sign inversion of $g_z$-component of 2D electron g-factor with narrowing of a hosting GaAs/Al$_{0.33}$Ga$_{0.67}$As quantum well (QW).  The energy splitting between spectral lines correspoding to recombination from the electron lowest Landau level and possessing opposite curcular polarization is analyzed. Due to exchange interaction, this value reveals \itshape maximum \normalfont or \itshape minimum \normalfont as a function of magnetic field strength in the vicinity of electron filling factor $\nu=3$. The actual type of extremum depends on the electron $g_z$-component sign. The method allows to detect the sign inversion of electron Lande g-factor with its absolute value changing of less than $\Delta g \approx 0.06$. The reversal happens at a QW width of ($65\pm 3$ \rm \AA).


\end{abstract}

\smallskip

\medskip

Величину расщепления уровней энергии носителей заряда в твердом теле, проявляющегося в магнитном поле (внешнем или внутреннем, например, ядерном) ввиду наличия у них спиновой степени свободы, принято описывать в терминах эффективного фактора Ланде, или g-фактора. Известно, что в результате спин-орбитального взаимодействия значение g-фактора электрона в полупроводнике может очень значительно отличаться от его величины $g_0=+2.0023$ для свободного электрона в вакууме ~\cite{Precise} и достигать больших отрицательных значений порядка $g_e\approx-50$ для узкозонного полупроводника InSb ~\cite{Roth}; при этом также типично проявляются зависимость фактора Ланде от магнитного поля ~\cite{AnisEPR} и его анизотропия в полупроводниковых системах с нарушенной симметрией ~\cite{AnisEPR, Anton, Kalevich}. 

В пионерской работе ~\cite{Roth} было проведено рассмотрение данного вопроса в теоретической модели, хорошо описывающей ряд объемных полупроводников IV группы и соединений III-V. Было получено аналитическое выражение для величины g-фактора электронов в зоне проводимости, в котором ведущими параметрами оказались ширина запрещенной зоны полупроводника и величина спин-орбитального расщепления между его двумя верхними валентными зонами. Таким образом, стала понятна перспектива создания объемных материалов с заданными значениями g-фактора электронов за счет, например, подбора композиции тройного сплава элементов III и V группы; данная возможность была наглядно продемонстрирована в работе ~\cite{Weis1} на примере соединений Al$_x$Ga$_{1-x}$As. 

С развитием технологий роста полупроводниковых гетероструктур, в частности, квантовых ям (КЯ), появился еще один способ управлением фактора Ланде электронов полупроводника, выступающего материалом КЯ ~\cite{InAs}, ~\cite{Snelling}. Размерное квантование в тонком слое GaAs приводит к модификации устройства валентной зоны (расщеплению ветвей легких и тяжелых дырок), изменению эффективных масс носителей заряда, но самый главный эффект для изменения величины g-фактора проистекает от увеличения "`эффективной ширины запрещенной зоны"', вызванного выталкиванием уровней размерного квантования электронов и дырок ~\cite{Ivch}. Таким образом, по мере уменьшения ширины КЯ наблюдается переход от объемного значения g-фактора в GaAs $g_e=-0.44$ через ноль (при ширине КЯ около 70 \rm \AA  ~\cite{Ivch})в положительные величины, отвечающие материалу барьера Al$_x$Ga$_{1-x}$As ($x>0.2$) ~\cite{Snelling} ~\cite{Yakovlev}; также появляется различие между его компонентой в направлении роста КЯ и компонентами, лежащими в её плоскости ~\cite{AnisEPR}.

Именно материалы с нулевым значением g-фактора электронов представляют особый интерес для сформулированных задач спинтроники ~\cite{SpinTr}, в которой манипуляции со спиновой степенью свободы частиц призваны дополнить или даже заменить подходы классической электроники, оперирующей зарядом электрона. Однако экспериментальные исследования достаточно узких КЯ GaAs/Al$_x$Ga$_{1-x}$As с близкой к нулю величиной фактора Ланде электронов проводились, как правило, с номинально нелегированными структурами методами оптической ориентации ~\cite{OptOrient}  ~\cite{Kalevich} либо наблюдения связанных с прецессией спина в магнитном поле квантовых биений в Керровском вращении ~\cite{Kerr},  эмиссии ~\cite{Emiss} или поглощении ~\cite{Abs} излучения. Транспортные методики, связанные с детектированием электронного парамагнитного резонанса ~\cite{EPR, AnisEPR, Anton} в проводящем канале двумерных электронов по разогреву электронной подсистемы, и наиболее точно определяющие величину спинового расщепления электронов, перестают работать в области малых величин g-фактора ввиду обстоятельств неоднородности образца и требования всё более низких температур эксперимента. По этой же причине оказывается неприменимым и подход, связанный с изучением неупругого (рамановского) рассеяния света на возбуждениях, связанных с переворотом спина ~\cite{Kulik}. Оптические эксперименты над заряженными КЯ по Керровскому вращению немногочисленны ~\cite{Larionov}, и вызывают вопросы об отношении получаемых результатов измерений для электронов, возбужденных светом над поверхностью Ферми, к заполненным состояниям "`темновых"' электронов. Таким образом, актуальная задача определения абсолютной величины и знака g-фактора электрона в GaAs легированной квантовой яме с $g_e\approx 0$, а также события переворота спина (то есть, детектирования электронного парамагнитного резонанса на "`нулевой частоте"'), оставалась по сути нерешенной.

Данная работа предлагает метод определения знака z-компоненты g-фактора (далее - просто g-фактора) электронов в GaAs легированной КЯ, находящихся на нижайшем уровне Ландау в сильном перпендикулярном магнитном поле вблизи фактора заполнения $\nu=3$. Оптическая спектроскопия фотолюминесценции и анализ циркулярно-поляризованных спектров в магнитном поле около 6 Тесла позволяют уверенно детектировать отличие в знаке электронного g-фактора для квантовых ям шириной 60 \rm \AA \rm и 70 \rm \AA, что соответствует изменению величины фактора Ланде порядка $\Delta g_e \approx 0.06$ .

\begin{figure}[h]\begin{center}
\includegraphics*[width=1.0\textwidth]{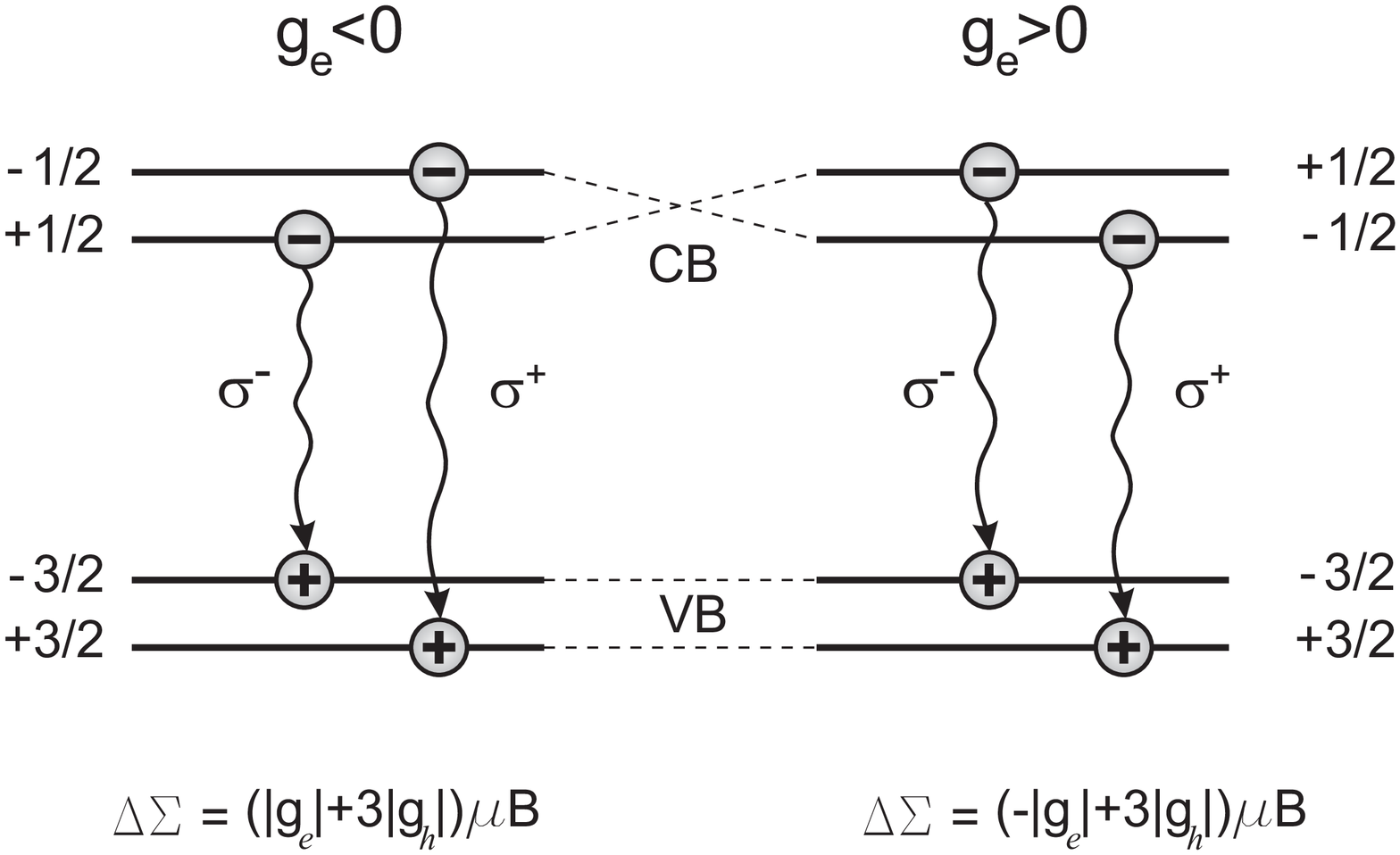}
\end{center}
\vspace{-2.mm} \caption{Схема оптических переходов в магнитном поле при рекомбинации электронов с нижайшего уровня Ландау и валентных дырок в GaAs квантовой яме при разных знаках электронного g-фактора.} \label{fig1} \vspace{-1.5mm}
\end{figure}

Хорошо известно, что при приложении магнитного поля вдоль направления роста легированной электронами GaAs квантовой ямы, её спектр фотолюминесценции расщепляется на компоненты, отвечающие рекомбинации с заполненных орбитальных уровней Ландау ~\cite{PL}. В достаточно сильных магнитных полях проявляется дальнейшее расщепление, связанное со спиновыми степенями свободы рекомбинирующих электрона и фотовозбужденной дырки. При этом анализ циркулярной поляризации в геометрии Фарадея показывает противоположные поляризации люминесценции с двух спиновых подуровней одного электронного уровня Ландау, а также определенное энергетическое расщепление $\Delta\Sigma$ между соответствующими спектральными линиями. В одночастичной модели рекомбинации единичного электрона и единичной дырки, связанных в экситон, данное расщепление определяется спиновыми расщеплениями электронного и дырочного уровней (в пренебрежении обменного взаимодействия). Рис.1 иллюстрирует схему нижайших по энергии оптических переходов в GaAs квантовой яме в магнитном поле в случае нелегированной КЯ, либо фактора заполнения электронов $\nu<2$, для ситуаций $g_e<0$ и $g_e>0$ ($|g_h|$, $\mu$ и $B$- дырочный g-фактор, магнетон Бора и величина магнитного поля соответстввенно). Видно, что в рамках данной модели величину электронного g-фактора в принципе можно было бы определить посредством измерения расщепления $\Delta\Sigma$, однако на практике данный подход оказывается неприменимым ввиду наличия дырочного вклада от второго слагаемого $3|g_h|$ в формулах на рис.1, величина которого для широких квантовых ям в слабых магнитных полях составляет порядка 1, а в общем случае заранее неизвестна ~\cite{Winkler}. Тем более затруднено детектирование малых изменений электронного g-фактора вблизи $g_e\approx 0$, когда происходит перестройка положения спиновых подуровней электрона.

\begin{figure}[h]\begin{center}
\includegraphics*[width=0.5\textwidth]{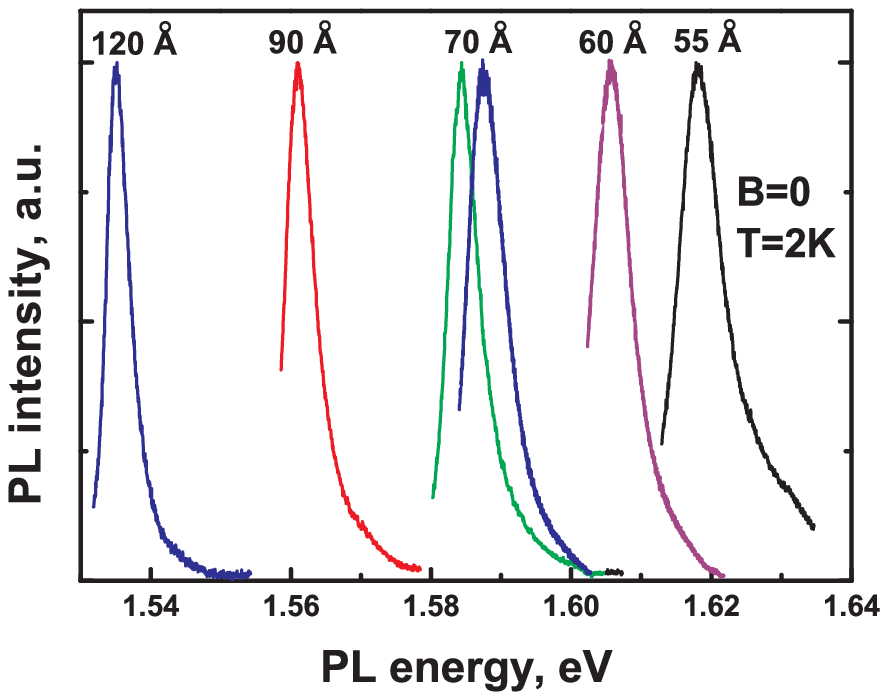}
\end{center}
\vspace{-2.mm} \caption{Спектры фотолюминесценции из изученных образцов КЯ в нулевом магнитном поле.
} \label{fig2} \vspace{-1.5mm}
\end{figure}

Ранее нами было показано ~\cite{JETP2010}, что описанная выше модель оптических переходов неприменима к описанию экспериментальных данных по расщеплению в поляризационно-разрешенных спектрах фотолюминесцеции из легированных GaAs КЯ при факторах заполнения $\nu>2$. Причина состоит в наличии обменных взаимодействий в коллективе двумерных электронов, в результате чего наблюдаемые энергетические расщепления проявляют осциллирующий с магнитным полем характер, и вблизи нечетных факторов заполнения имеют величину, существенно превосходящую суммарный вклад от "`голых"' электронных и дырочных спиновых расщеплений. Этот факт используется в данной работе для обнаружения смены знака электронного g-фактора вблизи $g_e\approx 0$.

Была исследована серия высококачественных образцов, содержащих одиночную легированную квантовую яму GaAs с номинальной шириной от 120 \rm \AA до 55 \rm \AA, заключенную в барьеры из Al$_{0.33}$Ga$_{0.67}$As. Гетеструктуры выращивались методом молекулярно-пучковой эпитаксии (MBE). В таблице \ref{table1} приведены основные характеристики исследованных образцов (темновая концентрация электронов, их транспортная подвижность, концентрация под светом, экстраполяция в B=0 результатов экспериментов ЭПР по определению z-компоненты g-фактора). Оптическое возбуждение и сбор сигнала фотолюминесценции осуществлялось с помощью двух световодов, источником излучения служил перестраиваемый титан-сапфировый лазер. Длина волны лазера выбиралась такой, чтобы осуществлять внутриямное фотовозбуждение с энергией фотона немного выше (25-30 мэВ) изучаемых оптических переходов; это позволяет избежать изменения концентрации электронов в яме при изменении магнитного поля. Эксперименты проводились в магнитном поле до 14 Тесла в геометрии Фарадея при температуре 2К. Спектры фотолюминесценции детектировались одиночным монохроматором с дисперсией 4 \rm \AA/mm , оснащенным охлаждаемым жидким азотом ПЗС-детектором. Для записи спектров в двух циркулярных поляризациях использовалась комбинация из фазовой пластинки $\lambda/4$ и линейного поляризатора, располагаемых непосредственно над поверхностью образца. Смена детектируемой поляризации осуществлялась посредством изменения направления магнитного поля в соленоиде.

\begin{center}

\begin{table}
\caption{\label{table1} Основные характеристики исследованных образцов.}

\begin{tabular}{|c|c|c|c|c|c|}
\hline
Sample & QW nominal, \AA & $n_{dark}$, $10^{11}\,cm^{-2}$ & $\mu$, $10^{3}\,\frac{cm^{2}}{Vs}$ & $n_{opt}$, $10^{11}\,cm^{-2}$ & $g_{zz}$ ESR data \\

\hline 82118&120&3.05&250&5.8&0.295\\

\hline 82119&90&2.7&65&5.9&0.182\\

\hline 82120&70&2.4&100&5.4& not detected \\

\hline 82121&55&1.8&25&5.1& not detected\\

\hline С1083&70&1.9&90&4.4& not detected\\

\hline С1084&60&1.7&90&4.2& not detected\\

\hline
\end{tabular}
\end{table}
\end{center}

На рис.2 приведены спектры фотолюминесценции двумерного электронного газа из всех образцов, снятые в нулевом магнитном поле. Несовпадение спектров для двух образцов с номинально одинаковыми КЯ 70 \rm \AA объясняется фактическим различием в их ширинах, составившем один монослой (примерно 3 \rm \AA). Для образцов с квантовыми ямами 120 и 90 \rm \AA  \rm протяженность спектральной линии определяется фермиевской энергией электронной системы, для более узких КЯ существенный вклад вносят ростовые флуктуации ширины.

\begin{figure}[h]\begin{center}
\includegraphics*[width=.5\textwidth]{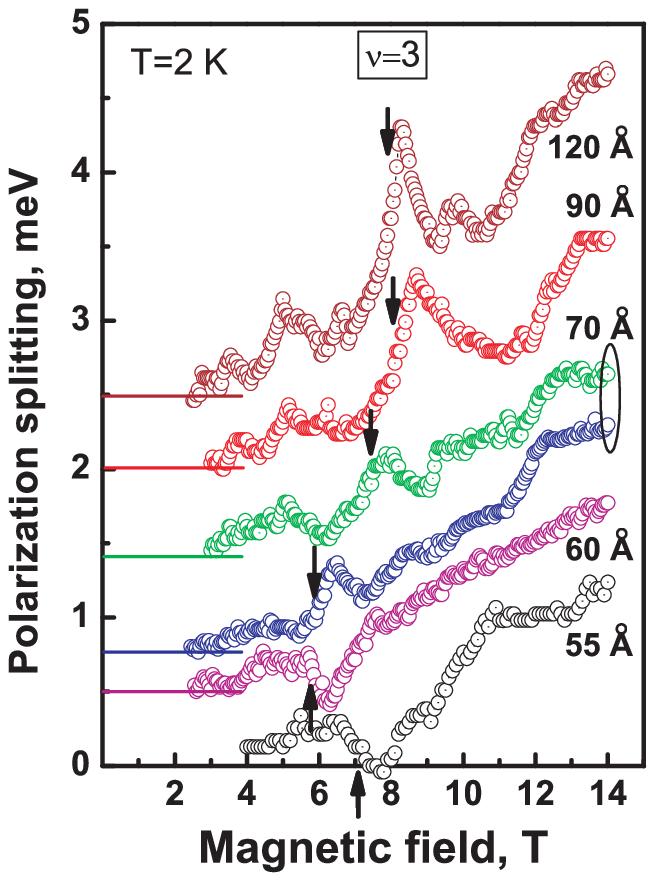}
\end{center}
\vspace{-2.mm} \caption{Магнитополевые зависимости энергетического расщепления между спектральными компонентами в $\sigma^{+}$ и $\sigma^{-}$ поляризациях, отвечающих рекомбинации с разных спиновых подуровней нижайшего уровня Ландау электронов. Для ясности рисунка, зависимости для разных образцов смещены по оси ординат; положение нулей указано соответствующими горизонтальными линиями. Стрелками отмечено значение магнитного поля, при котором реализуется фактор заполнения $\nu=3$  для каждого образца.
} \label{fig3} \vspace{-1.5mm}
\end{figure}

Магнитополевые зависимости энергетического расщепления между спектральными компонентами, отвечающими рекомбинации с двух спиновых подуровней нижнего электронного уровня Ландау, приведены со смещением по оси ординат на рис.3. Они демонстрируют характерное немонотонное поведение, однозначно связанное с фактором заполнения электронной системы (рис.4) и полуколичественно объясненное в работе ~\cite{JETP2010}. Особенностью исследованных здесь более узких КЯ является существенно меньшая величина электронной подвижности, вследствие чего выраженные осцилляции в расщеплении проявляются только в достаточно сильных магнитных полях более 4 Тесла. Как и в работе ~\cite{JETP2010}, для КЯ с ширинами 120, 90 и 70 \rm \AA наблюдается максимум в магнитполевой зависимости расщепления вблизи фактора заполнения $\nu=3$ (его положение для каждого образца отмечено стрелками на рис.3), когда величина спектрального расщепления существенно превышает суммарное расщепление электронного и дырочного спинового подуровня.

\begin{figure}[h]\begin{center}
\includegraphics*[width=1.0\textwidth]{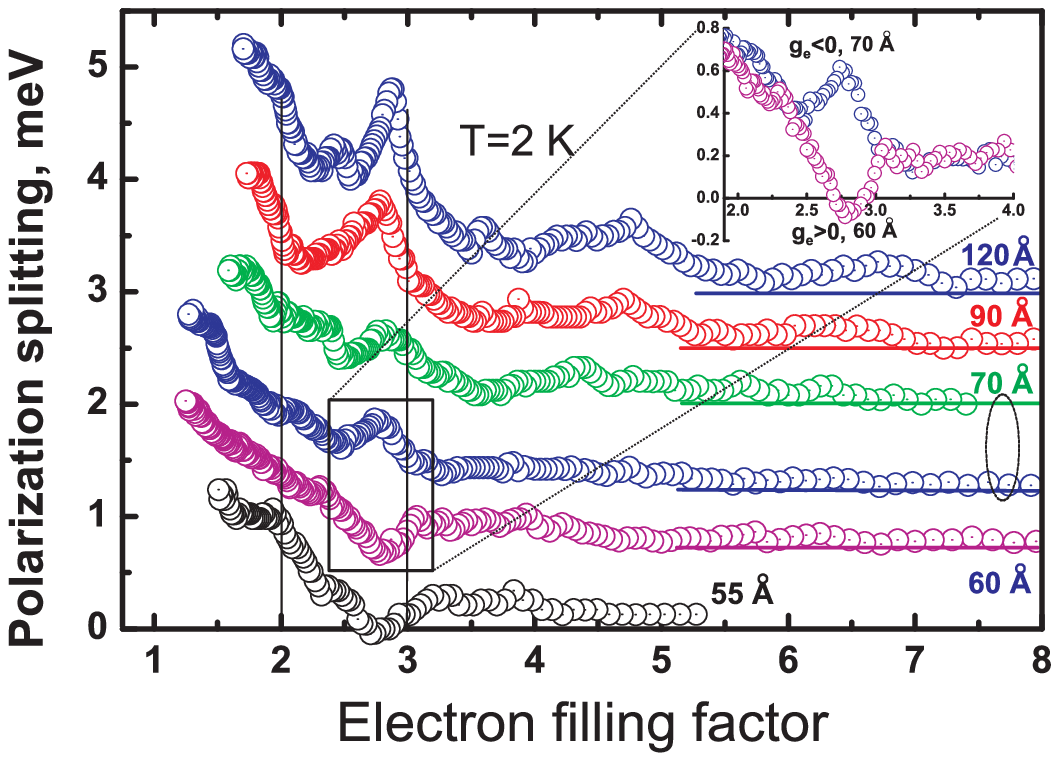}
\end{center}
\vspace{-2.mm} \caption{Зависимость энергетического расщепления из рис.3, построенная как функция электронного фактора заполнения. Изменение характера экстремума вблизи фактора заполнения $\nu=3$, более детально показанное на вставке (без смещения по оси ординат), отражает обращение знака электронного g-фактора при переходе от ширины КЯ 70 \rm \AA к 60 \rm \AA
} \label{fig4} \vspace{-1.5mm}
\end{figure}

Однако, при переходе к квантовым ямам шириной 60 и 55 \rm \AA, данная особенность изменяет характер: вместо локального \itshape максимума \normalfont наблюдается \itshape минимум \normalfont. Наиболее ярко это видно из вставки рис.4, где произведено сравнение двух несмещенных зависимостей спектрального расщепления для КЯ  70 и 60 \rm \AA. Изменение характера экстремума объясняется разными знаками электронного g-фактора в этих КЯ в магнитном поле около 6 Тесла. Действительно, обменно-усиленный вклад в энергетическое расщепление $\Delta\Sigma$ между спектральными компонентами с разной поляризацией (рис.1) всегда имеет тот же знак, что и вклад от "`голого"' электронного g-фактора, поэтому при изменении знака g-фактора и инверсии  электронных спиновых подуровней происходит обращение эффекта от этого вклада ~\cite{Note1}. 

Именно при значении ширины GaAs КЯ около 70 \rm \AA \rm было теоретически предсказано ~\cite{Ivch} нулевое значение электронного g-фактора. Отметим, что описанные эксперименты были проведены в достаточно сильном магнитном поле, которое вносит небольшую корректировку в величину g-фактора по сравнению с нулевым магнитным полем. К сожалению, предложенный подход не может предоставить количественную оценку величины фактора Ланде электронов, например, по масштабу наблюдаемого спектрального расщепления (так, на рис.3 оно практически одинаковое для КЯ шириной 120 и 90 \rm \AA, несмотря на различие в модуле g-фактора более в 1.6 раз, Таблица \ref{table1}). Однако, сравнение с расчетами работы ~\cite{Ivch} позволяет сделать вывод о том, что метод позволяет уверенно отличать знаки g-фактора при изменении его величины не более чем на $\Delta g_e \approx 0.06$. 

Таким образом, обнаружено, что z-компонента g-фактора электронов в GaAs/Al$_{0.33}$Ga$_{0.67}$As КЯ меняет знак при ширине ямы ($65\pm 3$ \rm \AA).

Работа была выполнена при частичной поддержке РФФИ.

\medskip

\bfseries Figure Captions: \normalfont

\bfseries РИСУНОК 1 \normalfont

Схема оптических переходов в магнитном поле при рекомбинации электронов с нижайшего уровня Ландау и валентных дырок в GaAs квантовой яме при разных знаках электронного g-фактора.

\bfseries РИСУНОК 2 \normalfont

Спектры фотолюминесценции из изученных образцов КЯ в нулевом магнитном поле.

\bfseries РИСУНОК 3 \normalfont

Магнитополевые зависимости энергетического расщепления между спектральными компонентами в $\sigma^{+}$ и $\sigma^{-}$ поляризациях, отвечающих рекомбинации с разных спиновых подуровней нижайшего уровня Ландау электронов. Для ясности рисунка, зависимости для разных образцов смещены по оси ординат; положение нулей указано соответствующими горизонтальными линиями. Стрелками отмечено значение магнитного поля, при котором реализуется фактор заполнения $\nu=3$  для каждого образца.

\bfseries РИСУНОК 4 \normalfont

Зависимость энергетического расщепления из рис.3, построенная как функция электронного фактора заполнения. Изменение характера экстремума вблизи фактора заполнения $\nu=3$, более детально показанное на вставке (без смещения по оси ординат), отражает обращение знака электронного g-фактора при переходе от ширины КЯ 70 \rm \AA \rm к 60 \rm \AA.


\begin{thebibliography}{2002}


\bibitem{Precise}
B. Odom, D. Hanneke, B. D’Urso, and G. Gabrielse, Phys. Rev. Lett. 97, 030801  (2006)

\bibitem{Roth}
L.M. Roth, B.Lax and S.Zwerdling,  Phys. Rev. 114, 90 (1959)

\bibitem{AnisEPR}
Yu. A. Nefyodov, A. V. Shchepetilnikov, I. V. Kukushkin, W. Dietsche, and S. Schmult, Phys. Rev. B 83, 041307(R) (2011)

\bibitem{Anton}
A. V. Shchepetilnikov, Yu. A. Nefyodov, I. V. Kukushkin and W. Dietsche, Journal of Physics: Conference Series 456, 012035 (2013)

\bibitem{Kalevich}
В.К. Калевич, В.Л. Коренев, Письма в ЖЭТФ 56, 257 (1992)




\bibitem{Weis1}
C.Weisbuch, C. Hermann, Phys. Rev. B 15, 816 (1977)



\bibitem{InAs}
T. P. Smith III and F. F. Fang, Phys. Rev. B 35, 7729 (1987)

\bibitem{Snelling}
M. J. Snelling, G. P. Flinn, A. S. Plaut et al., Phys. Rev. B 44, 11345 (1991)

\bibitem{Ivch}
Е.Л.Ивченко, А.А.Киселев, ФТП 26 (8), 1471 (1992)

\bibitem{Yakovlev}
I. A. Yugova, A. Greilich, D. R. Yakovlev et al., Phys. Rev. B 75, 245302 (2007)

\bibitem{SpinTr}
S. A. Wolf, D. D. Awschalom, R. A. Buhrman et al. , Science 294, 1488 (2001)

\bibitem{OptOrient}

M. J. Snelling, G. P. Flinn, A. S. Plaut, R. T. Harley, A. C. Tropper, R. Eccleston, and C. C. Phillips, Phys. Rev. B 44, 11345 (1991)



\bibitem{Kerr}

I. A. Yugova, A. Greilich, D. R. Yakovlev et al. , Phys. Rev. B 75, 245302 (2007)

\bibitem{Emiss}
A. P. Heberle, W. W. Ruhle, and K. Ploog, Phys. Rev. Lett.72, 3887 (1994)

\bibitem{Abs}

S. Bar-Ad and I. Bar-Joseph, Phys. Rev. Lett.66, 2491 (1991)

\bibitem{EPR}

D. Stein, K. v. Klitzing, and G. Weimann Phys. Rev. Lett. 51, 130 (1983); M. Dobers, K. v. Klitzing, and G. Weimann, Phys. Rev. B 38, 5453 (1988)

\bibitem{Kulik}

Л.В. Кулик, В.Е. Кирпичев, УФН 176, 365–382 (2006) ; L. V. Kulik, S. Dickmann, I. K. Drozdov, A. S. Zhuravlev, V. E. Kirpichev, I. V. Kukushkin, S. Schmult, W. Dietsche, Phys Rev B 79 , 121310 (2009)

\bibitem{Larionov}
А. В. Ларионов, А. С. Журавлев, Письма в ЖЭТФ т. 97 (3), 156 (2013)

\bibitem{PL}
I. V. Kukushkin, V. B. Timofeev, Adv.Physics 45, 147 (1996)

\bibitem{Winkler}

"`Spin-orbit Coupling Effects in Two-Dimensional Electron and Hole Systems"' by Roland Winkler, Springer Tracts in Modern Physics Vol. 191, Springer, Berlin (2003)

\bibitem{JETP2010}
В. В. Соловьев, И. В. Кукушкин, Ш. Шмульт, Письма в ЖЭТФ т.92 (9) 665 (2010) (V.V. Solov'ev , I.V. Kukushkin and S.Schmult, JETP Letters , 92(9), 600 (2010))


\bibitem{Note1}

Предполагается, что знак g-фактора валентных дырок остается неизменным. Это предположение подтверждается теоретическим расчетом (см. ~\cite{Winkler}) для исследованных в настоящей работе ширин квантовых ям и магнитных полей более 4 Тесла.

\end{thebibliography}
\end{document}